\documentclass[11pt,a4paper,useAMS]{emulateapj}

\usepackage{natbib}
\usepackage{amsmath}
\usepackage{graphicx}
\usepackage{morefloats}

\def\nicefrac#1#2{
    \raise.5ex\hbox{#1}%
    \kern-.1em/\kern-.15em%
    \lower.25ex\hbox{#2}}
\def\ltapprox{\raise 2pt \hbox {$<$} \kern-1.1em \lower 5pt \hbox {$\approx$}}
\def\ltsim{\raise 2pt \hbox {$<$} \kern-1.1em \lower 4pt \hbox {$\sim$}}
\def\gtsim{\raise 2pt \hbox {$>$} \kern-1.1em \lower 4pt \hbox {$\sim$}}
\def\arcsec{$^{\prime\prime}\,$}
\def\arcmin{$^{\prime}\,$}

\def\ie{{\it i.e.,~}}
\def\eg{{\it e.g.,~}}
\def\gtsim{\; \raise0.3ex\hbox{$>$\kern-0.75em \raise-1.1ex\hbox{$\sim$}}\; }
\def\ltsim{\; \raise0.3ex\hbox{$<$\kern-0.75em \raise-1.1ex\hbox{$\sim$}}\; }

\begin{document}


\title{LOFAR discovery of a radio halo in the high-redshift  galaxy cluster PSZ2 G099.86+58.45}


\author{
R.~Cassano\altaffilmark{1},  A.~Botteon\altaffilmark{1,2,3}, G. Di Gennaro\altaffilmark{3}, 
G. Brunetti\altaffilmark{1}, M. Sereno\altaffilmark{4,5}, T.W. Shimwell \altaffilmark{6,3}, R.J. van Weeren\altaffilmark{3}, M. Br\"uggen\altaffilmark{7}, F. Gastaldello\altaffilmark{8}, L. Izzo\altaffilmark{9}, L. B\^{i}rzan\altaffilmark{7}, A. Bonafede\altaffilmark{2,7}, V. Cuciti\altaffilmark{7}, F. de Gasperin\altaffilmark{7}, H.J.A. R\"ottgering\altaffilmark{3}, 
M. Hardcastle\altaffilmark{10}, 
A.P. Mechev\altaffilmark{3},
C. Tasse\altaffilmark{11,12}}


\affil{\altaffilmark{1}INAF - Istituto di Radioastronomia, via P. Gobetti 101, I--40129 Bologna, Italy}
\affil{\altaffilmark{2} Dipartimento di Fisica e Astronomia, Universit\`{a} di Bologna, via P. Gobetti 93/2, I-40129 Bologna, Italy}
\affil{\altaffilmark{3} Leiden Observatory, Leiden University, P.O. Box 9513, 2300 RA Leiden, The Netherlands}
\affil{\altaffilmark{4} INAF - Osservatorio di Astrofisica e Scienza dello Spazio, via P. Gobetti 93/3, I-40129 Bologna, Italy}
\affil{\altaffilmark{5}INFN, Sezione di Bologna, viale Berti Pichat 6/2, 40127 Bologna, Italy}
\affil{\altaffilmark{6} ASTRON - Netherlands Institute for Radio Astronomy, PO Box 2, 7990 AA Dwingeloo, The Netherlands}
\affil{\altaffilmark{7} Universit\"at Hamburg, Hamburger Sternwarte, Gojenbergsweg 112, 21029, Hamburg, Germany}
\affil{\altaffilmark{8} INAF/IASF - Milano, Via A. Corti 12, I--20133 Milan, Italy}
\affil{\altaffilmark{9}Instituto de Astrof\'isica de Andaluc\'ia (IAA-CSIC), Glorieta de la Astronom\'ia s/n, E-18008, Granada, Spain}
\affil{\altaffilmark{10} Centre for Astrophysics Research, School of Physics, Astronomy and Mathematics, University of Hertfordshire, College Lane, Hatfield AL10 9AB, UK}
\affil{\altaffilmark{11} GEPI, Observatoire de Paris, Universit\'e PSL, CNRS, 5 Place Jules Janssen, 92190, Meudon, France}
\affil{\altaffilmark{12} Department of Physics \& Electronics, Rhodes University, PO Box 94, Grahamstown, 6140, South Africa}

\email{E-mail: rcassano@ira.inaf.it}


\shorttitle{Diffuse radio emission in PSZ2 G099.86+58.45}
\shortauthors{}

\vspace{0.5cm}
\begin{abstract}
\noindent 
In this Letter, we report the discovery of a radio halo in the high-redshift galaxy cluster PSZ2 G099.86+58.45 ($z=0.616$) with the LOw Frequency ARray (LOFAR) at 120-168 MHz. This is one of  the most distant radio halos discovered so far. The diffuse emission extends over $\sim$ 1 Mpc and has a morphology similar to that of the X-ray emission as revealed by \textit{XMM-Newton} data. The halo is very faint at higher frequencies and is barely detected by follow-up 1-2 GHz Karl G.~Jansky Very Large Array (JVLA) observations, which enable us to constrain the radio spectral index to be $\alpha\ltsim 1.5-1.6$, \ie with properties between canonical and ultra-steep spectrum radio halos.
Radio halos are currently explained as synchrotron radiation from relativistic electrons that are re-accelerated in the intra-cluster medium (ICM) by turbulence driven by energetic mergers. We show that in such a framework radio halos are expected to be relatively common at $\sim150$ MHz ($\sim30-60\%$) in clusters with mass and redshift similar to PSZ2 G099.86+58.45; however, at least 2/3 of these radio halos should have steep spectrum and thus be very faint above $\sim 1$ GHz frequencies. 
Furthermore, since the luminosity of radio halos at high redshift depends strongly on the magnetic field strength in the hosting clusters, future LOFAR observations will also provide vital information on the origin and amplification of magnetic fields in galaxy clusters.

\vspace{4mm}
\end{abstract}



\section{Introduction}

Cluster-scale ($\sim$Mpc-scale), diffuse synchrotron emission is frequently found in high-mass ($M_{500}\gtsim5\times10^{14}$ M$_{\odot}$) merging galaxy clusters in the form of so-called giant radio halos (hereafter RHs), apparently unpolarised emission that fills large cluster volumes
\citep[\eg][]{2019SSRv..215...16V}.
These sources are characterised by low surface brightnesses and steep spectra  ($\alpha >1$, with $S_{\nu}\propto \nu^{-\alpha}$, this is the convention we adopt in this paper). Their properties and connection with clusters mergers support the idea that they trace turbulent regions in the intra-cluster medium (ICM) where relativistic particles are trapped and re-accelerated during cluster-cluster mergers \citep[\eg][]{2014IJMPD..2330007B}.

Present statistical studies are limited to relatively low-redshift systems \citep[$z\ltsim 0.35-0.4$; \eg][]{2007A&A...463..937V,2008A&A...484..327V,2015A&A...579A..92K,2015A&A...580A..97C}, with only a handful of halos discovered at $z\simeq0.5-0.6$ \citep{2000NewA....5..335G,2009A&A...503..707B,2012MNRAS.426...40B,2009A&A...505..991V} and only two at higher redshift, one located in the \emph{El Gordo} galaxy cluster, an exceptionally massive object at $z=0.87$ \citep{2014ApJ...786...49L} and one in PLCKG$147.32-16.59$, a cluster at $z=0.65$ \citep{2014ApJ...781L..32V}. 
At high redshift an increasing fraction of the energy that is dumped into the acceleration of relativistic electrons in RHs is radiated away in the form of inverse Compton (IC) emission, $dE/dt \propto (1+z)^4$. This is expected to cause a decline of the fraction of RHs in high-$z$ galaxy clusters with respect to their low-$z$ counterparts \citep{2006MNRAS.369.1577C}.
In particular the fraction of clusters hosting RHs at higher $z$ is also expected to be very sensitive to the magnetic fields in these systems.

The discovery of high-redshift RHs has been limited by the need for radio observations with high sensitivity to steep spectrum emission
and high resolution to adequately distinguish the emission from contaminating sources. The advent of
the LOw Frequency ARray \citep[LOFAR;][]{2013A&A...556A...2V}, which can produce deep, high-resolution, high-fidelity, low-frequency radio images, has opened up the possibility to study RHs at low frequencies with unprecedented detail and sensitivity \citep[\eg][]{2016ApJ...818..204V,2016MNRAS.459..277S,
2018MNRAS.478.2927B,2018MNRAS.478..885B,2019A&A...622A..19B,2018MNRAS.478.2218H,2019A&A...622A..20H,2019A&A...622A..21H,2018MNRAS.473.3536W,2019A&A...622A..25W}.

In this Letter, we report on the discovery of a radio  
halo in the high-redshift galaxy cluster PSZ2 G099.86+58.45 (PSZ2G099, hereafter) which was observed as part of the LOFAR Two-metre Sky Survey (LoTSS).
LoTSS is an ongoing sensitive $\sim 100\mu$Jy/beam, high-resolution, $\sim 6''$, 120-168 MHz survey of the entire northern sky \citep{2017A&A...598A.104S,2019A&A...622A...1S}.

PSZ2G099 is a massive, $M_{500}=(6.84\pm0.48)\times 10^{14}$ M$_{\odot}$ \citep{2016A&A...594A..27P} and hot, $kT=8.9^{+2.8}_{-1.1}$ keV \citep{2018NatAs...2..744S} cluster discovered through its Sunyaev-Zel'dovich (SZ) signal by the {\it Planck} satellite. Recently, the gravitational lensing signal of this cluster has been traced up to 30 Mpc from its centre implying that it sits in a very high-density environment, about six times denser than the average $\Lambda$CDM prediction at this redshift \citep{2018NatAs...2..744S}. The expected complex dynamics around this cluster make the discovery of a RH in this cluster particularly interesting.

 Hereafter, we adopt a $\Lambda$CDM cosmology with $H_{0} = 70$~km~s$^{-1}$~Mpc$^{-1}$, $\Omega_{m} = 0.3$, and $\Omega_{\Lambda} = 0.7$. With the adopted cosmology, 1\arcsec~corresponds to a length scale of 6.766~kpc at $z=0.616$.  

\begin{figure}[t]
\begin{center}
\includegraphics[trim =0cm 0cm 0cm 0cm,width=0.5\textwidth]{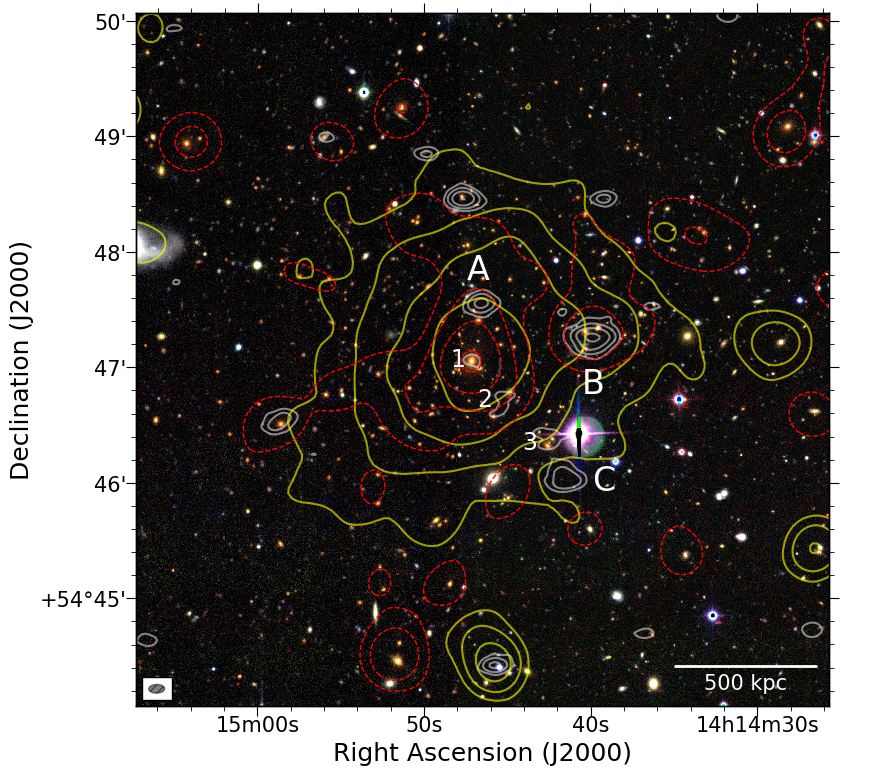}
\end{center}
\caption{CFHTLenS $g$, $r$ and $i$ band composite images of PSZ2G099 with overlaid the LOFAR 144 MHz high resolution (8.3\arcsec $\times$ 4.3\arcsec) contours (white), the \textit{XMM-Newton} contours (yellow) and light density contours of the cluster member galaxies (red-dashed). Contours are spaced by a factor of 3 starting from $5\sigma$ for LOFAR (where $\sigma=70$ $\mu$Jy beam$^{-1}$) and by a factor of 2 starting from $5.0 \times 10^{-6}$ counts s$^{-1}$ pixel$^{-1}$ for \textit{XMM-Newton} (cf. Fig.~\ref{fig:lofar}, center). The LOFAR beam is shown in the lower left corner. 
Labels and numbers show the position of the radio sources (see Sect.\ref{sec:results} for details).}
\label{fig:optical}
\end{figure}

\section{Observations \& data reduction}
\label{sec:obs}

\begin{figure*}[ht]
\begin{center}
\includegraphics[width=.33\textwidth,trim={0.8cm 0cm 0.8cm 0cm},clip]{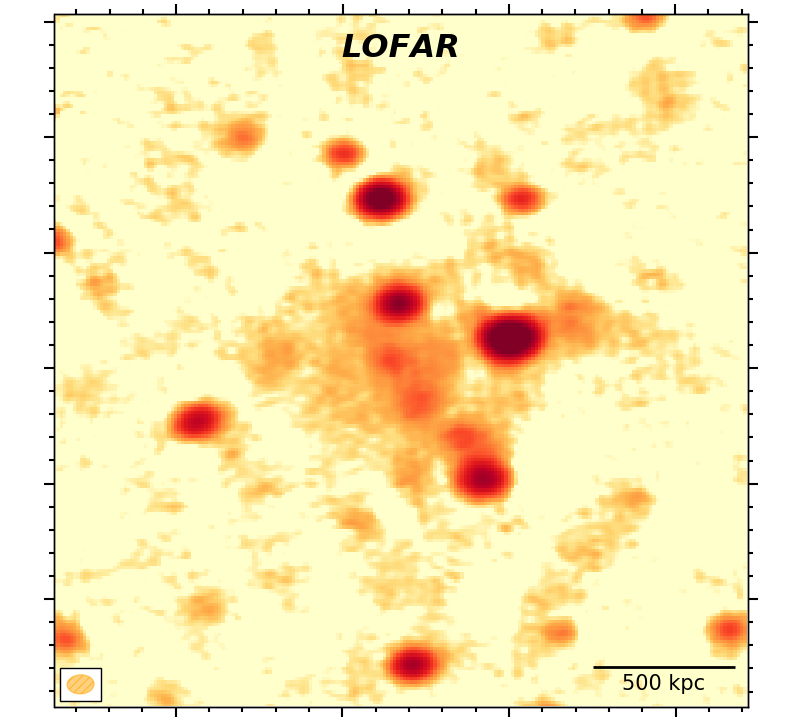}
\includegraphics[width=.33\textwidth,trim={0.8cm 0cm 0.8cm 0cm},clip]{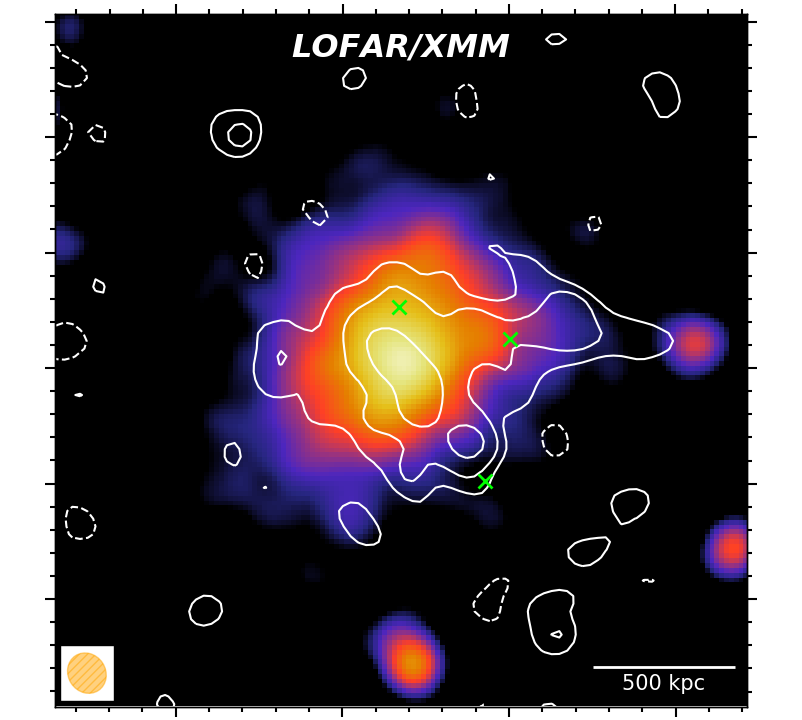}
\includegraphics[width=.33\textwidth,trim={0.8cm 0cm 0.8cm 0cm},clip]{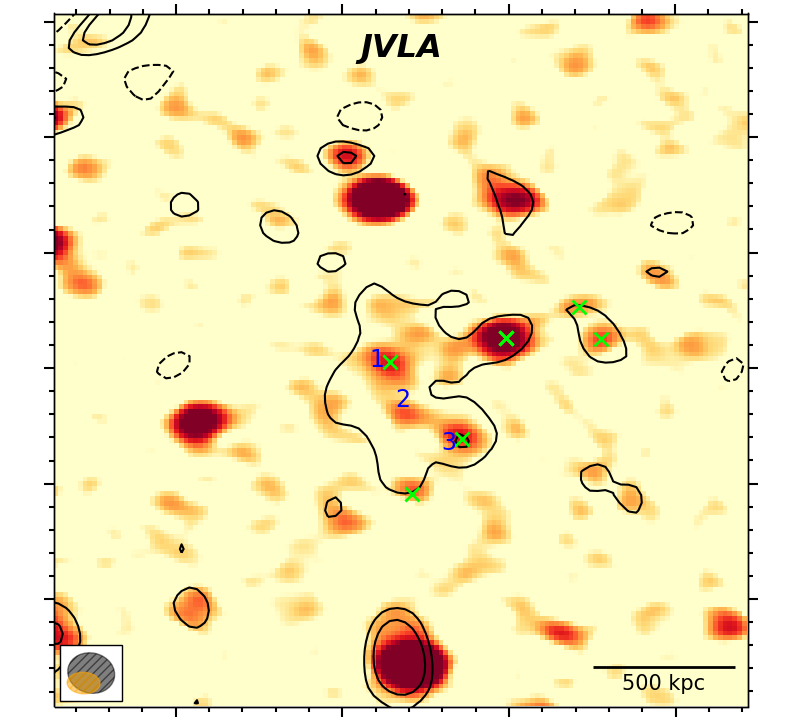}
\end{center}
\caption{Left: LOFAR 144 MHz medium-resolution (13.9\arcsec $\times$ 9.7\arcsec, rms noise $\sigma=90$ $\mu$Jy beam$^{-1}$). Center: \textit{XMM-Newton} image in the $0.5-2.0$ keV band smoothed with a 3-pixel Gaussian kernel (1 pixel = 2.5\arcsec) with low-resolution source-subtracted LOFAR contours (21.6\arcsec $\times$ 19.4\arcsec, $\sigma=200$ $\mu$Jy beam$^{-1}$) spaced by a factor of 2 starting from 2$\sigma$. In both panels, $-2\sigma$ contours are displayed in dashed lines while the beam shapes are shown in the bottom left corners. Right: 1--2 GHz high-resolution (17.1\arcsec $\times$ 11.2\arcsec, $\sigma=$ $20 \mu$Jy beam$^{-1}$) JVLA image with $2.5\sigma$ and $5\sigma$ low-resolution source-subtracted JVLA contours (24.5\arcsec $\times$ 20.8\arcsec, $\sigma=36$ $\mu$Jy beam$^{-1}$). Numbers in the right panel show the position of the three blobs also highlighted in Fig.~\ref{fig:optical}.}
\label{fig:lofar}
\end{figure*}

The LoTSS pointings consist of 8~hr observations in the $120-168$ MHz band, which are typically separated by $\sim 2.6^\circ$.
 PSZ2G099 is located at $\sim 15$\arcmin from the center of pointing P$214+55$ (in the region of the HETDEX Spring Field), acquired on 14 May 2015 (ObsID L343224).
Data reduction of this pointing was performed with the pipeline described by \cite{2019A&A...622A...1S}, which performs direction-independent and dependent calibration and imaging of the full LOFAR field-of-view, using \textsc{prefactor} \citep{2016ApJS..223....2V,2016MNRAS.460.2385W,2019A&A...622A...5D,2018A&C....24..117M}, \textsc{killMS} \citep{2014A&A...566A.127T,2014arXiv1410.8706T,2015MNRAS.449.2668S} and \textsc{DDFacet} \citep{2018A&A...611A..87T}. To improve the image quality in the direction of PSZ2G099, we used the products of the pipeline, subtracted all the sources outside a region of 15\arcmin $\times$ 15\arcmin surrounding the target, and performed extra phase and amplitude self-calibration loops in this sub-field (more details by van Weeren et al. in prep.). The LOFAR images shown in this paper were produced with WSClean v2.6 \citep{2014MNRAS.444..606O} and have a central frequency of 144 MHz.
Uncertainties on the LOFAR integrated flux densities are dominated by errors in the absolute flux scale, which is conservatively set to 20$\%$, in line with LoTSS measurements \citep{2019A&A...622A...1S}.

A follow-up observation of PSZ2G099 was carried-out with the Karl G. Jansky Very Large Array (JVLA) in the L-band (\ie 1--2 GHz) in C- and D-configurations, for a total time of 4 hrs. Following the procedure described by \cite{2018ApJ...865...24D}, we calibrated the antenna delays, bandpass, cross-hand delays, and polarisation leakage and angles using the primary calibrators 3C286 and 3C147. The calibration solutions were then applied to the target, and self-calibration on the single dataset was performed to refine its amplitude and phase solutions. During the final self-calibration on the combined dataset we also employed an additional self-calibration on a bright source located at the edge of the primary beam, whose side lobes affect the cluster radio emission. All the images in this paper were produced with \texttt{CASA v5.0} \citep{2007ASPC..376..127M} using w-projection (Cornwell et al. 2005, 2008), Briggs weighting with \texttt{robust=0}, \texttt{nterms=3} (Rau \& Cornwell 2011) and are corrected for the primary beam attenuation.

PSZ2G099 was observed three times with \textit{XMM-Newton} (ObsID: 0693660601, 0693662701, 0723780301) for a total exposure time of 63~ks. Periods of the observations that were affected by soft proton flares are excluded during the analysis by using the Scientific Analysis System tasks. The displayed image is a background-subtracted and exposure-corrected mosaic image in the $0.5-2.0$ keV band of the three ObsIDs where MOS and pn camera images are combined.

\begin{table}[t]
\begin{center}
\caption{Radio halo properties}
\begin{tabular}{ll}
\hline
\hline
$S_{\rm{halo,\,144\:MHz}}$  (mJy)$^{a}$ & $25.3\pm5.7$ \\
$S_{\rm{halo,\,1500\:MHz}}$  (mJy) & $\sim 1.5\pm0.5$ \\
$P_{\rm{halo,\,1.4\:GHz}}$ ($10^{24}$ W Hz$^{-1}$)$^{b}$& $2.85 \pm 0.95$\\
Halo Size (Mpc) & 0.6-1.2 \\
144~MHz compact source fluxes (mJy) & $7.0 \pm 1.4$ (A)  \\
& $30.9\pm 6.2$ (B) \\
& $7.4 \pm 1.5$ (C)\\
\hline
\hline
\end{tabular}
\label{tab:properties}
\end{center}
$^{a}$ {integrated (within $2\sigma_{\rm{rms}}$) RH flux density.}\\
$^{b}$ {assuming a spectral index of $\alpha=1.2$ for the halo emission, $k$-corrected.}\\\
\end{table}

\section{Results}
\label{sec:results}

A multi-wavelength view of PSZ2G099 is reported in Fig.~\ref{fig:optical} where the LOFAR high-resolution contours are overlaid on the CFHTLenS \citep[Canada France Hawaii Telescope Lensing Survey,][]{2012MNRAS.427..146H} image highlighting the presence of the compact radio sources in the direction of the cluster. The X-ray contours from {\it XMM-Newton} are also reported to show the distribution of the thermal gas. 
The red contours follow the average surface density of the cluster member galaxies selected in the optical $i$-band. We select 3058 galaxies from the CFHTLenS image with photometric redshift within $\pm0.06(1+z_\text{cl})$ of the cluster redshift, comparable to the survey photo-$z$ uncertainty. The distribution was smoothed with a Gaussian kernel with a standard deviation of $50~\text{kpc}/h$. 
The double-peaked average surface density and the elongated X-ray emission indicate that PSZ2G099 is in a merging phase.

At medium resolution (Fig.~\ref{fig:lofar}, left panel), LOFAR clearly reveals extended diffuse emission at the center of the galaxy cluster. Subtracting the emission from point sources and tapering down to $\sim20$\arcsec~resolution (central panel) we find a total extent of $\sim3\times1.5$\arcmin in the east-west and north-south direction, respectively, corresponding to a physical extent of $\sim1.2\times0.6~$Mpc. We classify this emission as a RH due to its extension, morphology, and location in the cluster.
The integrated flux density of the sources labelled A, B, and C located in the RH region (Fig.~\ref{fig:optical}) is reported in Tab.~\ref{tab:properties}. Among these only source B is detected in the JVLA image and has a spectral index $\alpha\simeq 1.5$ (see Fig.~\ref{fig:lofar}, right panel), meaning that source A and C are very steep spectrum sources ($\alpha>1.8$).
In the high-resolution LOFAR image, another three blobs of emission are detected at $5\sigma$ ( numbered from 1 to 3 in Fig.~\ref{fig:optical}) which account for a small total flux density of $S_{\rm{144\:MHz}} \sim 3$ mJy. These are also detected at high-frequency, as seen in Fig.~\ref{fig:lofar} (right panel).

To disentangle the contribution of the point sources and provide a reliable measurement of the RH flux density we create an image of diffuse emission only (Fig.~\ref{fig:lofar}, central panel) by subtracting the clean components of the sources A, B and C from the visibilities obtained by applying an inner \textit{uv}-cut of 3.5 k$\lambda$ (corresponding to a linear size of about 400 kpc at $z=0.616$) to the data.
The flux density of the halo measured within the LOFAR $2\sigma$ contour is $S_{\rm{144\:MHz}} = 25.3 \pm 5.7$ mJy (this excludes the $3$ mJy flux of the three blobs). We verified that the subtraction of the three point sources using \textit{uv}-cuts in the range $2.0-5.5$ k$\lambda$ (corresponding to scales of $\sim700-250$ kpc at the cluster redshift) corresponds to variations of the RH flux density of $\sim27.2-31.6$ mJy, which is within the calibration error. The diffuse flux density is $\sim15\%$ lower if emission above the $3\sigma$ level is considered. 

In Fig.~\ref{fig:lofar} (right panel) we show the JVLA high-resolution (17.1\arcsec $\times$ 11.2\arcsec) image with the 2.5$\sigma$ contours from the JVLA low-resolution source-subtracted image. The latter image has been obtained after the subtraction of the clean components of the point sources in the field (including those outside the cluster region) using an \textit{uv}-cut of 1.3 k$\lambda$ (corresponding to $\sim500$ kpc) and \texttt{robust=-0.5}, and than tapering at $\sim$20\arcsec resolution. The rms noise of this map is $36~\mu$Jy/beam. Diffuse emission from the RH region is barely detected (the halo is not clearly detected at 3$\sigma$) but shows some similarity with the RH seen by LOFAR. However, the JVLA image appears to be contaminated by residuals of radio emission by individual sources.
Indeed, from the un-tapered map in which point sources have been subtracted, we estimate that a residual contribution of $\sim 1$ mJy can still be attributed to point sources which leads to a residual RH flux density of $1.5$ mJy within $2\sigma$.
Because of the very low brightness of the halo we evaluated the statistics of the diffuse flux extracted from similar regions around the RH measuring residual flux densities which range between $\approx -0.3$ mJy and $+0.6$ mJy. This provides an uncertainty in the determination of the RH flux of $\approx\pm 0.5$ mJy at $1.5$ GHz.

\begin{figure}[t]
\includegraphics[width=0.47\textwidth]{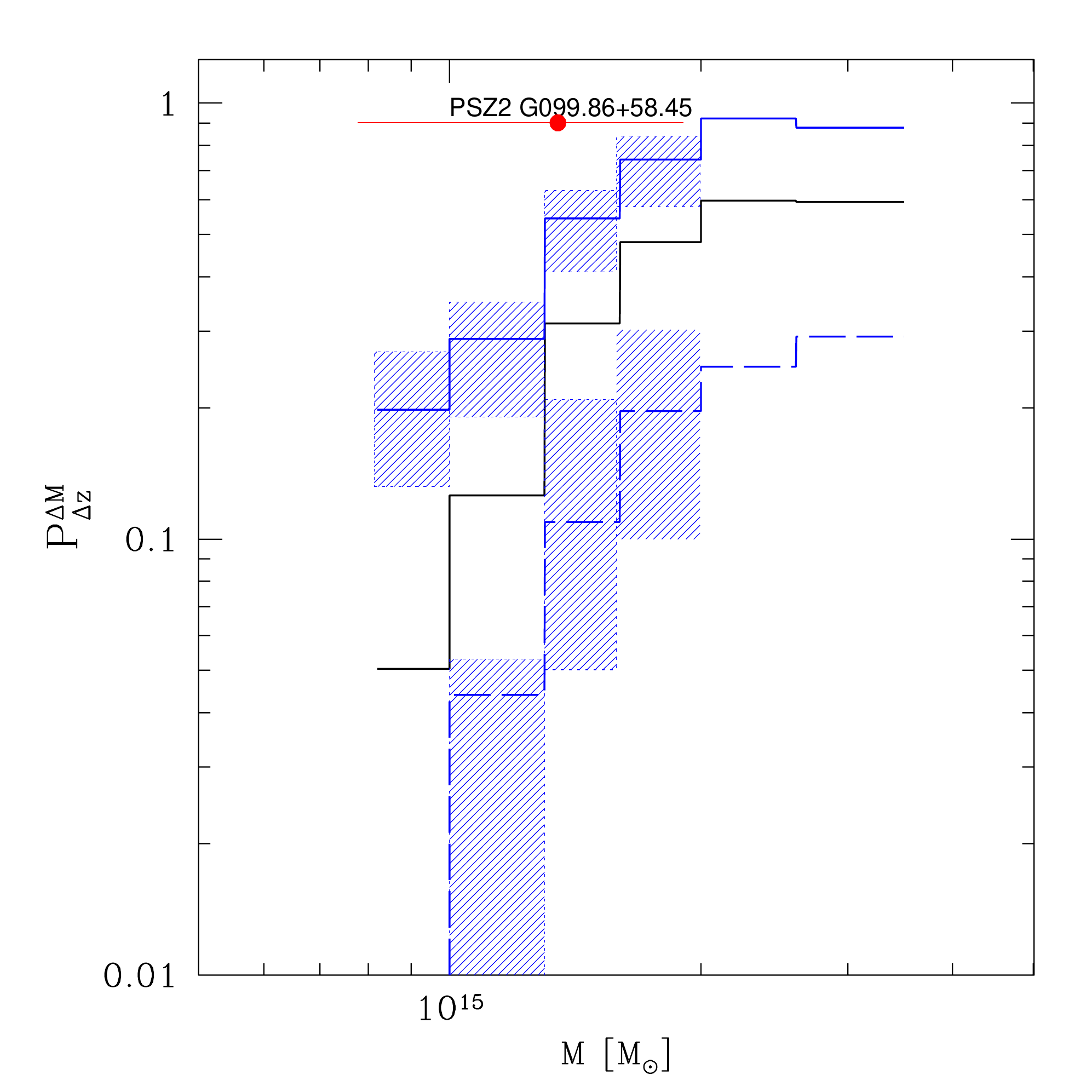}
\caption{Probability to form RHs with $\nu_s\gtsim 120$ MHz (solid curves) and with $\nu_s\gtsim 1400$ MHz (dashed curves) as a function of the cluster virial mass in the redshift range $0.6-0.7$, for 
$B= 1~\mu$G (black curve) and $B=4.8~\mu$G (blue lines). For the latter case, the 1$\sigma$ uncertainty derived through Monte Carlo calculations is also shown (blue shadowed regions, see text).
}
\label{fig:theory}
\end{figure}
 
\vspace{1cm}
 \section{Discussion}
\label{sec:discussion}

Observing high-$z$ RHs provides unique information on the physics of these sources in an extreme environment. In fact, 
 models for the origin of RHs at such early times are challenged by the strong IC-losses of relativistic electrons ($dE/dt \propto E^2(1+z)^4$) which would compete and hamper the acceleration of high-energy electrons and reduce the synchrotron luminosity that is generated by a factor $B^2/B_{cmb}^2$.
Under these conditions the maximum synchrotron frequency emitted by the electrons accelerated in RHs depends on $B/B_{cmb}^2$ \citep{2006MNRAS.369.1577C} and thus the magnetic field sets the frequency window where the emission can be observed.

To investigate this point, we derive the formation probability of RHs in a cluster with mass and redshift similar to PSZ2G099 using the statistical model developed by \cite
[][see also Cassano et al. 2006, 2010]{2005MNRAS.357.1313C} which is based on the turbulent re-acceleration scenario. 
These models match the observed fraction of clusters with RHs derived in mass-selected cluster sample at low-$z$ \citep{2015A&A...580A..97C}.  
In such a framework, the synchrotron spectra of RHs steepen at high frequencies since turbulent re-acceleration is balanced by radiative losses of relativistic electrons.
This steepening makes it difficult to detect RHs at frequencies higher than the frequency $\nu_s$ at which the steepening becomes severe\footnote{$\nu_s$ is defined as the frequency where the synchrotron spectrum is $\alpha=1.9$, with $\alpha$ calculated between $\nu_s/2.5$ and $\nu_s$ \citep[see][for details]{2010A&A...509A..68C}}. 
As a simplified approach to estimate the occurrence of RHs at a given frequency $\nu_0$ we assume that only halos with $\nu_s>\nu_0$ can be observed at $\nu_0$.
Massive, merging clusters should statistically have larger values of $\nu_s$ making them detectable at relatively high frequencies ($\sim$ GHz), while less energetic mergers are expected to produce RHs with lower $\nu_s$ which are only detectable at low frequencies ($<$ few 100 MHz). The latter are referred to as ultra-steep spectrum RHs (USSRHs\footnote{RHs with $\alpha>1.5$ are considered USSRHs (see discussion in Brunetti 2004; Brunetti et al. 2008) }) and are predicted to be the dominant class of RHs in low-frequency radio surveys, such as LoTSS \citep{2008Natur.455..944B,2010A&A...509A..68C,2018MNRAS.473.3536W}. Theoretically, the two classes of RHs, the canonical and USSRHs, mark the extremes of a continuous distribution of properties of these radio sources.

In Fig.~\ref{fig:theory}, we show the probability to form RHs with $\nu_s\gtsim 120$ MHz (solid curves) and with $\nu_s\gtsim 1400$ MHz (dashed curves) for clusters in the redshift range $0.6-0.7$. We assume two values for the average magnetic field strength in the RH region:$\sim 1~\mu$G (black line) and $B=4.8~\mu$G (blue line). The latter value corresponds to the magnetic field that maximises the lifetime of relativistic electrons radiating at a given observing frequency at the redshift of the system (\ie $B\simeq B_{cmb}/\sqrt{3}\simeq 4.8 ~\mu$G),
and thus it should be considered an upper limit for the model expectations.
For the case $B=4.8~\mu$G, we also report the resulting probabilities derived by 1000 Monte Carlo extractions of galaxy cluster samples from the pool of simulated merger trees (shadowed blue regions). Considering the virial mass of PSZ2G099, $M_v\simeq (1.35\pm 0.57)\times10^{15}\,M_{\odot}$ \citep{2017MNRAS.472.1946S}, we find that at low frequency the probability to form a RH in such a cluster is maximum 40 to 60$\%$,
while it drops to about $30\%$ considering $B\sim 1~\mu$G. We conclude that RHs in clusters similar to PSZ2G099 should be fairly common in LOFAR surveys. On the other hand, the probability to form RHs in these systems at high frequencies ($\nu_s\simeq1400$ MHz) is found to depend critically on the magnetic field in the emitting volume. The maximum probability ($B=4.8~\mu$G) is about $20\%$ whereas a very small probability (below the percent level, not visible in Fig.~\ref{fig:theory}) is obtained considering $B=1~\mu$G. The drop in probability between low and high frequencies is essentially due to the presence of USSRHs that glows up preferentially at lower frequencies.

For the specific case of PSZ2G099 the spectrum is not measured well, however our constraint $\alpha\ltsim1.5-1.6$ implies that we can exclude the case of a RH with an extremely steep spectrum,
and that the radio properties of PSZ2G099 could be intermediate between those of canonical and USSRHs.
We measure a RH flux density at 1.5 GHz of $\sim 1.5\pm 0.5$ mJy, implying a 1.4~GHz radio power ($k$-corrected) $P_{1.4}\simeq (2.85\pm0.95)\times 10^{24}$ W Hz$^{-1}$, which is consistent with the $P_{1.4}-M_{500}$ correlation observed in intermediate redshift clusters (Cassano et al. 2013).
This provides information on the magnetic field in the RH volume. The expected radio luminosity is $P_{syn}\propto \eta_{rel}(\rho v_t^3/L_{inj})/(1+(B_{cmb}/B)^2)$ (where $\eta_{rel}$ account for the fraction of turbulence dissipated in particle acceleration). Although the turbulent dissipation rate ($\rho v_t^3/L_{inj}$, with $\rho$ gas density, $v_t$ turbulent velocity and $L_{inj}$ the turbulence injection scale) can be larger in dynamically-young high-$z$ clusters than in low-$z$ ones, the fact that the radio power of PSZ2G099 is similar to its lower-$z$ counterparts suggests that $B$ in this cluster is at least similar to that in low-$z$ ones.

This finding provides important information for models of amplification of $B$ in galaxy clusters, in particular on the origin of the seed field that is stirred and amplified by turbulence and small scale dynamo \citep{2005JCAP...01..009D,2018MNRAS.474.1672V,2018SSRv..214..122D}. Since the dynamo mechanism is a slow process that requires several turbulent eddy-turnover times \citep[several Gyrs, \eg][and ref. therein]{2018MNRAS.474.1672V} our observations suggest an important role of Active Galactic Nuclei (AGN) and Galactic Winds (GW) in setting a significant seed field in the ICM at high redshift.

At the same time it should be stressed that PSZ2G099 is sitting in a special region of the Universe, which was found to be about six times denser than the average density of the Universe at that redshift \citep{2018NatAs...2..744S}; its RH could therefore be unique. The effect of the large-scale environment \citep[environment-bias; see][and ref. therein]{2018MNRAS.474.5143M} can trigger the formation of a RH: in such a place more merger/accretion episodes are expected to {\it bias} the halo growth with respect to cluster of similar mass in less denser regions of the Universe. As a consequence, future analysis of LOFAR surveys will be very important to constrain the formation rate of RHs at high-$z$ and the origin of magnetic fields in these systems.

\section{Conclusions}
\label{sec:conclusions}

We report on the discovery of a $\sim$~Mpc {(total extent)} RH in the PSZ2G099 cluster using LOFAR observations at $120-168$ MHz carried out for the LoTSS. Being at a redshift of $\sim0.616$ it is among the most distant RHs discovered so far, and the furthest away currently discovered by LOFAR. 

The halo is also barely detected by 1-2 GHz follow-up JVLA observations that constrains the radio spectral index to be $\alpha\ltsim 1.5-1.6$, \ie with properties between canonical and ultra-steep spectrum radio halos. The estimated 1.4 GHz radio power locates the halo on the observed $P_{1.4}-M_{500}$ correlation meaning that the magnetic field strength in this cluster should be not much different from that of other low to intermediate redshift halos in clusters with similar mass.

We show that current turbulent reacceleration models predict that RHs in clusters similar to PSZ2G099 should be common ($\sim30-60\%$) at low frequencies, however at least 2/3 of these RHs should be USSRHs and thus be very faint at high-frequencies.
This shows the power of LOFAR as a unique machine to discover RHs at high-$z$. 

Furthermore, since the fraction of clusters with RHs (and with USSRHs) at high-$z$ and their luminosity depend on the magnetic field in these systems we claim that LOFAR statistical studies of high-z RHs will provide vital information on the origin of magnetic fields in galaxy clusters.

\acknowledgments
{\it Acknowledgments:}
This paper is based on data obtained from the International LOFAR Telescope (ILT) under project code LC3\_008. LOFAR \citep{2013A&A...556A...2V} is the Low Frequency Array designed and constructed by ASTRON. It has observing, data processing, and data storage facilities in several countries, which are owned by various parties (each with their own funding sources), and are collectively operated by the ILT foundation under a joint scientific policy. The ILT resources have benefitted from the following recent major funding sources: CNRS-INSU, Observatoire de Paris and Universit\'{e} d'Orl\'{e}ans, France; BMBF, MIWF-NRW, MPG, Germany; Science Foundation Ireland (SFI), Department of Business, Enterprise and Innovation (DBEI), Ireland; NWO, The Netherlands; The Science and Technology Facilities Council, UK; Ministry of Science and Higher Education, Poland; Istituto Nazionale di Astrofisica (INAF), Italy. This research made use of the Dutch national e-infrastructure with support of the SURF Cooperative (e-infra 180169) and the LOFAR e-infra group. The J\"{u}lich LOFAR Long Term Archive and the German LOFAR network are both coordinated and operated by the J\"{u}lich Supercomputing Centre (JSC), and computing resources on the Supercomputer JUWELS at JSC were provided by the Gauss Centre for Supercomputing e.V. (grant CHTB00) through the John von Neumann Institute for Computing (NIC). This research made use of the University of Hertfordshire high-performance computing facility and the LOFAR-UK computing facility located at the University of Hertfordshire and supported by STFC [ST/P000096/1], and of the LOFAR IT computing infrastructure supported and operated by INAF, and by the Physics Dept. of Turin University (under the agreement with Consorzio Interuniversitario per la Fisica Spaziale) at the C3S Supercomputing Centre, Italy.
M.S. acknowledges financial contribution from ASI-INAF n.2017-14-H.0. The Leiden LOFAR team acknowledge support from the European Research Council under the  FP/2007-2013/ERC Advanced Grant NEWCLUSTERS-321271. RJvW acknowledges support from the VIDI research programme, project number 639.042.729, which is financed by the NWO. ABon acknowledges financial support from the ERC-Stg DRANOEL, no 714245, and from the MIUR grant FARE SMS. L.I. acknowledges support from funding associated with Juan de la Cierva Incorporacion fellowship IJCI-2016-30940. APM acknowledge support from the NWO/DOME/IBM programme ``Big Bang Big Data: Innovating ICT as a Driver For Astronomy'', project \#628.002.001.
We would like to thank the anonymous referee for useful comments that help to improve the presentation of our work.



{\it Facilities:} \facility{LOFAR}, \facility{XMM-Newton}, \facility{JVLA}

\end{document}